\documentstyle[osa,manuscript]{revtex}  

\begin{document}                

\title{Molecular Dynamics Simulations of a Pressure-induced Glass Transition}

\author{Shelly L. Shumway, Andrew S. Clarke$^*$ and Hannes J\'onsson }
\address{Department of Chemistry, BG-10, University of Washington,
Seattle, Washington 98195}
\maketitle
\begin{abstract}
We simulate the compression of
a two-component Lennard-Jones liquid at a variety of constant
temperatures using a molecular dynamics algorithm  in an isobaric-isothermal
ensemble.    The viscosity of the liquid increases with pressure,  undergoing
a broadened transition into a structurally arrested,  amorphous state.
This transition, like the more familiar one induced by cooling,
is correlated with a significant increase
in icosahedral ordering.   In fact, the structure of the final
state, as measured by an analysis of the bonding,
is essentially the same in the
glassy, frozen state whether
produced by squeezing or by cooling under pressure.
We have computed an effective
hard-sphere packing fraction at the
transition, defining the transition pressure or temperature
by a cutoff in the diffusion constant,
analogous to the traditional laboratory definition  of the glass
transition  by an arbitrary, low cutoff in viscosity.
The packing fraction at this transition point is not constant, but
is consistently higher for runs
compressed at higher temperature.
We show that this is because the transition point defined by
a constant cutoff in the diffusion constant is not the same as the
point of structural arrest, at which further changes in
pressure induce no further structural changes, but that
the two alternate descriptions may be reconciled by using
a thermally activated cutoff for the diffusion constant.    This
enables estimation of the characteristic activation energy for diffusion at
the point of structural arrest.

\end{abstract}

\section{ Introduction}
  The glass transition most often studied is that which occurs when
a viscous liquid forms a glass
upon reduction of the temperature at constant, usually zero,
pressure.  A
transition into a solid amorphous structure could alternately be
obtained at constant temperature by applying pressure.
We examine this latter transition here, through molecular
dynamics [MD] simulations of a two-component Lennard-Jones
[LJ] system.

The combination of pressure and temperature effects results
in behavior not seen in cooling studies without pressure.
We are particularly interested in the relationship between
structural and dynamical properties, and we address a non-trivial
aspect of this relationship in this paper, demonstrating that
diffusion, a dynamical
property, is correlated in an activated, non-linear
way with the structural changes that occur as the material is
compressed.

\section{Method of Simulation}
We, like many others who have studied glassy behavior through computer
simulations,\cite{1,jonsson}
have used a two-component Lennard-Jones [LJ] system as the
model:  interactions between atoms are described by the pair potential
$v_{\alpha \beta} = 4 \epsilon [(\sigma_{\alpha \beta}/r)^{12} -
(\sigma_{\alpha \beta}/r)^6].$  All data reported here were obtained for
a 2000-atom system consisting of 1600 small atoms with $\sigma_{11} = 1$
and 400  larger atoms with $\sigma_{22} = 1.2,$  with $\sigma_{12} = 1.1$ for
the interaction between the two different types.  The other parameters,
mass $m$ and energy $\epsilon,$ were the same for all atoms.   We used periodic
boundary conditions in all three directions.  The time step for
integration was $0.01 \tau,$ where
$\tau = ( m \sigma_{11}^2 / \epsilon)^{(1/2)}.$

We have simulated with a molecular dynamics algorithm in an isobaric-isothermal
ensemble, using Andersen's constant pressure method\cite{HCA}
combined with stochastic
collisions.  A detailed description of
this  algorithm  is given
in the appendices to a paper by Fox and Andersen.\cite{FA}
Andersen's method adds an
extra degree of freedom, the total volume, to the Lagrangian for the
system.  The volume adjusts freely to maintain
a nearly constant, though fluctuating,  pressure.   We show in Appendix
A how to calculate a reasonable value for the ``piston mass,''  which
determines the coupling strength of the volume term.
The stochastic collisions
involve effectively colliding a few randomly selected
particles each time step with
a heat bath that has a Boltzmann distribution of velocities at the correct
temperature. Andersen
has demonstrated\cite{HCA} that
the combination of his constant pressure method with
such stochastic collisions in standard molecular dynamics results in a
correct isothermal-isobaric ensemble:
the time-averaged value $<F>$ for any quantity F  converges to the correct
isobaric-isothermal ensemble average $F_{NPT}$ in the limits of
$t \rightarrow \infty$ and $N \rightarrow \infty.$

We have run at five different constant temperatures, in each case with
pressure
increasing over time as shown in Figure 1.  At a temperature of
$T^* = kT/\epsilon = 0.8,$  we have examined the behavior at
several intermediate pressures more closely by running for a long time at
constant pressure
with initial configurations given by the endpoints
of the plateaus in the rapidly-increasing-pressure runs, as shown in
the inset to Figure 1.
Pressure is given throughout this paper in
units of $\epsilon/\sigma_{11}^3,$  which would be
about 425 bars if the smaller particles were argon.

\section{ Diffusion}
The self-diffusion constant
$D_{\alpha} = ({d / dt} <\!{x_\alpha}^2\!>)/6$
is a convenient measure of how freely atoms of type $\alpha$
are able to move around within the system.
As the liquid becomes more
viscous and relaxations freeze out, atoms become stuck in their
places and the diffusion constant falls.
This is seen in Figure 2 for an average diffusion constant\cite{average}
$D = (D_1 + D_2)/2$ for each
of the constant temperatures we simulated.
At a temperature of
$T^* = 0.6,$ the diffusion
constant falls to nearly zero by the rather low pressure of  6,
while at a higher temperature of
$T^* = 1.5,$ diffusion persists to much higher pressures.
We have fit the data for the three lower temperatures to an exponential
function $D = A e^{-BP},$ for $T^* = 1.0$ to a sum of two exponentials,
$D = A e^{-BP} + C e^{-EP},$    and for $T^* = 1.5$ to a sum of three
exponentials. Our least-squares fitting program
returned $\chi^2 < 4 \times 10^{-5}$  in each case. The single
exponential was not used for the
two highest temperatures because the fit was bad, with
$\chi^2$ two orders of magnitude higher for $T^* = 1.0$ and four
for  $T^* = 1.5.$

Experimentally,
the glass transition in a real material is
defined by  an arbitrary
cutoff in viscosity ($\eta$): by
$\eta \sim 10^{14}$ poise, the liquid is
acting solid on human time scales.  Since the
behavior of the diffusion coefficient D is similar\cite{acw,anderson}
to that of $1/\eta$ in
glassy materials,  the transition could alternately be defined in
terms of an arbitrary cutoff in D;   $\eta$ is easier to measure
experimentally, D is easier in computer simulations.
We define our
glass transition pressure $P_g,$  which is plotted in Figure 3,
as the point at which
D falls below 0.001,  in units of $\sigma_{11}^2/\tau,$
using the fitting functions described above to
determine this point.
Some other
cutoff could equally well have been chosen; the behavior of the
$P_g$ versus $T$ curve is qualitatively the same for cutoffs ranging
from 0.0005 to
0.004, as shown in the inset to Figure 3.
Of course, any diffusion measured in our simulations
corresponds to a much lower viscosity
than $10^{14}$ poise.  Our cutoff of $D = 0.001$ corresponds to about
$5 \times 10^{-7}$ $\rm cm^2/s$ using argon units, or a diffusion length
of about $.4 \; \AA$ for the 40 ps constant pressure
plateaus of the fast runs,  or $2 \; \AA$ for the 1 ns extended plateaus
of the runs shown in the inset to Figure 1.  A liquid
in the laboratory with such a diffusion constant would still be
described as a viscous liquid,\cite{acw}  not a glass;  it is basically frozen
on time scales of our simulation,  though not
yet on laboratory time scales.

We have plotted the potential energy as a function of cube root of volume
in Figure 4, indicating the point corresponding to $P_g$
for each temperature.   At the beginning of each run, when the pressure
is low, thermal fluctuations keep the atoms farther apart on average
than would be
optimal energetically.  As the pressure goes up, atoms move closer
together,  mapping out a picture of the average potential seen by the
ensemble at each density.
Clarke\cite{clarke} noticed such an effect many years ago
in simulations of a one-component LJ system, and thought the
glass transition would generally tend to occur near the minimum
in the potential energy.  While this is indeed true at
some temperatures, it is not universal.  At
lower temperatures,
diffusion halts before the average spacing
even reaches the optimal value, almost as soon as thermal vibrations
start forcing atoms into the region dominated by the repulsive potential,
while at
higher temperatures, diffusion continues well into the repulsive
region.

\section{Icosahedral Ordering}

It has been previously observed in simulations and experiments in which
viscous liquids were cooled into a glassy state that an increase in
icosahedral bonding is often associated with vitrification.\cite{jonsson,JA}
An increase in local icosahedral ordering has also been observed
during densification of hard sphere packings.\cite{CJ}
This effect is also seen in the pressure-induced transition, as we now
show.

We have analyzed the bonding using a common-neighbor analysis\cite{FJ} [CNA]
in which a set of three indices $jkl$ specifies the local environment of
a bonded pair of atoms.  Two atoms are considered bonded if they are within
a certain distance of each other,  which depends on the type of each.
The cutoff is chosen here as the location of the first
minimum in the appropriate partial radial distribution function g(r).
The first index $j$ is the number of common neighbors, or the
number of atoms bonded to both.
The next index $k$ is the total number of bonds between these common
neighbors, and $l$ is the number of bonds in the longest continuous chain
formed by the $k$ bonds between the common neighbors.  In an FCC crystal,
for example, all bonds would be of the type ``421.''  Icosahedral bonding
results in ``555'' pairs, as illustrated in Figure 5.  Chemical ordering
is not reflected in this scheme; the label does not reflect the atomic
types of the bonded neighbors.

Because thermal vibrations obscure the underlying order, we use quenched
configurations for the structure analysis:
a steepest descent minimization of the potential energy under constant
pressure is performed until the nearest local minimum on the potential
energy surface is reached.
We then decompress the quenched
configurations, at zero temperature,
in order to bring everything to the same state for comparison,
as illustrated in Figure 6a.  Virtually no structural change is observed upon
repeated compression and  decompression at zero temperature, as shown in
Figure 6b, although this certainly would not be true at any finite temperature.
By bringing everything to the same state for comparison by a route in which
activated changes do not occur, we can see the
underlying structural changes resulting from activated processes
that took place during compression at finite temperatures.
The differences we see this way involve energy barriers,  and
no simple relaxation could take the system from one state to the other.

The nature of the bonding changes as the system rearranges in response to
pressure applied at finite temperature.
FCC and HCP bonding, identified as 421 and 422 pairs, decrease gradually,
and 433 pairs increase gradually, as shown in Figure 7a.
The only dramatic change is a strong increase in icosahedral bonding,
or 555 pairs, throughout the transition region,  stabilizing
at a maximum value at about the same pressure as the diffusion
constant becomes nearly zero (Figure 7b).

\section{ Isothermal Compressibility}

We evaluated the isothermal compressibility,
$\kappa =  -{1 \over V} (\partial V/\partial p)_T,$ throughout the transition.
For all temperatures, $\kappa$ was a smoothly varying function of pressure,
with nothing special at or near $P_g$.  As indicated
in Figure 8 for the $T^*=0.8$ run, $\kappa$ correlates well with the
diffusion constant. Both fall together as the pressure increases, although
compressibility never goes to zero, but continues to decrease
gradually after diffusion is halted. This is because further compression
continues to  push atoms closer together, further  into
the repulsive region, but without inducing the structural changes that
caused $\kappa$ to decrease so rapidly with pressure while the material was
still liquid.  Quenched, decompressed
configurations show a 1\% increase in density as a result of the structural
changes that take place as pressure increases up to the point of structural
arrest, so some anomaly in $\kappa$ would be expected in the transition
region. No such anomaly is seen,
although a small signal could easily be obscured by
the underlying continuous decrease in
$\kappa$ combined with noise in the data.

\section{WCA radius and Packing Fraction}

Hudson and Andersen\cite{HA} reported in 1978  that
vitrification in binary alloys tends to occur
at about the same effective hard-sphere packing fraction,
computed using an extension of
the method of Weeks, Chandler and Andersen\cite{WCA} [WCA]
for
several different materials,  and that computer models of hard spheres and
LJ fluids performed before that date all gave similar results.  The value
of this packing fraction, $\eta_g = \pi \rho R^3/6$, was
$0.53  \pm 0.02$.   They concluded that the
underlying physical process driving the transition is
dominated by the repulsive  part
of the interaction and the simple inability of atoms
at this density to slide freely past each other, whatever the details of
the interaction potential.

In the years since then many authors have
computed the packing fraction
at vitrification for various model systems,  and have generally concurred with
the concept introduced by Hudson and Andersen, although not everyone has
obtained
the same numbers for the packing fraction, perhaps due partly to differences
in the definition of the transition point  and
in the method for computing the effective radius.
For example, Clarke\cite{clarke}
found  vitrification
in a one-component Lennard-Jones system at a packing  fraction of 0.60, but
commented in a later paper\cite{acw} with Angell and Woodcock that his results
were
actually consistent with those of Hudson and Andersen  if the same criterion
for $\rm T_g$
were applied to both sets of data.  Alexanian and
Haywood\cite{alexanian_haywood} obtained
values of .586, .569, and .571 for a hard-sphere system in which the transition
point was determined respectively by a signal in the heat capacity, vanishing
of the diffusion coefficient, and  a calculated equation of state.  Berg and
coworkers\cite{berg}
obtained experimental values from fits to viscosity data of .537 for a binary
mixture of alkali metals and .547 for a ternary mixture.  Pusey and van
Megen\cite{PvM} found that for particles in colloidal suspension, it is .56.
Simulations by Cape and Woodcock\cite{cape_woodcock} and by
Ullo and Yip\cite{ullo_yip}
gave between .52 and .55 for soft sphere and truncated LJ systems.
Abraham\cite{abraham} found  values of .534-.55 for a one-component LJ system
using Monte Carlo simulations in which he, like us, studied the effects of
squeezing as well as of cooling, and he came to the conclusion that the packing
fraction at vitrification is the same for either squeezing or cooling.
Bengtzelius\cite{bengtzelius} performed theoretical
calculations for an LJ fluid and calculated that the effective packing fraction
would be .536-.552 at the glass transition, defined as the point at which
discontinuities appear in the specific heat and isothermal compressibility.

Following this large and illustrious group, we
have also computed an effective WCA packing fraction at vitrification,
defined here as the point at which
the diffusion coefficient falls below a cutoff
of 0.001.   We are interested in comparing glasses with different temperature
and pressure histories using the same criterion for vitrification and
exactly the same method for calculating the hard-sphere radius for
each, in order to identify any trends that may exist.

We used the WCA  method
with the corrections of Verlet and Weiss,\cite{VW} as outlined in Appendix
B,  to compute effective hard-sphere
radii $R$ near vitrification
for each of the five temperatures
we simulated, as well as for cooling runs at two different  pressures,
P=0 and P=8, for comparison.
$R$ is computed analytically
based on the number density $\rho$, temperature, and the form of the
interaction potential. The packing fraction $\eta$ is then computed as
$\pi \rho R^3/6.$  One subtlety in applying this formalism to
a two-component system is that the effective size of the two
components does not scale precisely as $\sigma^3;$ the large
atoms are a bit softer than small atoms, and are thus squeezed
more, so they occupy a bit less space.  Results reported
in this paper are calculated with the approximation
that size does scale with $\sigma^3,$
so that the packing fraction  is the same for different species.
This assumption is consistent with the work of
Lee and Levesque\cite{LL} in their
analysis of two-component Lennard-Jones systems.  Just to be sure, though,
we recalculated using  the location of the first peak in
$g(r)$ rather than $\sigma$ to determine relative atomic
sizes and thus effective density separately for each component.
The final packing fractions $\eta$ thus obtained agreed to within 2\%
in every case.

Our results for $\eta_g,$
the effective packing fraction
at the glass transition,  are given in Table 1.
For the zero-pressure cooling run, we obtain a value
$\eta_g = .564,$ in reasonable agreement with
values obtained by others.   Glasses prepared at high
pressures in the laboratory, according to Woodcock, Angell and
Cheeseman,\cite{wood_ang_cheese} tend to have higher densities
when examined at normal pressure than the same glasses
prepared at normal pressure.   Thus, it is not surprising that
our values for $\eta_g$ for the $P=8$ cooling run and also
for the squeezing runs are higher than those obtained at
zero pressure.   The reader might be surprised that the
packing fraction at $P_g$ for the $T^* = 1.5$ run exceeds
the hard sphere dense random packing limit of about 0.64,
but this limit does not apply here because  this is not
really a hard sphere system;  the WCA radius corresponds
closely to the position of the first peak in g(r),  but some
interatomic distances are as much as 12\% less than this value,
allowing densities that considerably exceed the corresponding
hard-sphere limit.    What we find remarkable about the results
is the trend in $\eta_g$ for the squeezing runs:
the higher the temperature at which the material
is compressed, the higher the packing fraction at which diffusion
falls below the cutoff.

\section{DISCUSSION}
This trend requires an explanation.   It is conceivable that
the material is able to densify better when compressed at higher
temperature, but we found no evidence for this.
Bonding, as determined by a CNA analysis, is the same within
error bars for all the structurally arrested final states
produced by compression at various temperatures, and also for the
$P=8$ cooling run.
Details of the structure are different in random ways
for every run, of course, but the average
occurrence of each type of bonded pair is the same within
our error bars.
Also, when the final
states are quenched and then brought to zero pressure, systematic differences
in density disappear for the squeezing runs and for the
$P=8$ cooling run.
(We should note that the P=0 cooling runs  were different:  they
produced glassy  states that are intrinsically less
dense, and with slightly different bonding.)
The large systematic increase in $\eta_g$ with higher temperature for
compression is thus not associated with any intrinsic structural
differences that we could detect in the glassy states produced.

There is a more subtle structural difference however: as the material
densifies during squeezing, the first part of the split second peak of g(r)
increases
in height, and when g(r) is scaled by the location of the first peak, the
second and further peaks move closer, as illustrated in Figure 9.
We have found a small, but consistent, trend
in g(r) scaled by the location of the first peak at the glass transition:
for higher temperature compression runs,  the  second,
third and fourth peaks are located at smaller values of r.
These values continue to decrease somewhat past the pressure at
which diffusion falls below the cutoff, and then stabilize.

A structural measure of the transition point, then, could be defined
as the pressure $P_g$ for which
further compression results in no change in the scaled g(r) function
because no further structural changes can occur.
  Another manifestation of structural arrest is that
the volume after quenching and then releasing the pressure becomes
constant.  (Quenching eliminates the temperature-dependent thermal
expansion, and releasing the pressure eliminates the overall rescaling
of atomic positions, allowing direct comparison of states with different
histories.)  This is plotted for our $T^*=0.8$ run in Figure 10.
Another correlated property is the disappearance of
anharmonicity: the only change the structurally arrested state
undergoes when quenched is the elimination of thermal noise, or
gaussian fluctuations of the atoms about their average positions.  The
rms value $(<x-x_0>^2)^{1/2}$ of these fluctuations is
proportional to temperature,
so when the harmonic system is quenched and atomic locations
before and after the quench are compared,  the average rms
difference is proportional to temperature, and independent
of pressure history or anything else.  This is certainly
not the case for the liquid state, in which quenching induces
considerable non-activated structural changes.
Rms differences
for the quenched versus unquenched configuration of the viscous fluid
decrease with pressure until structural arrest
is reached, at which point they settle down to a constant, as shown
in Figure 10.

We have measured these three properties for the $T^*=0.8$ run,
and have found that all three give
results  consistent with  the conclusion of
structural arrest occurring at a pressure of 14.
This is higher than the pressure of 10.5 at which diffusion falls
below the cutoff of 0.001; diffusion essentially ceased before
pressure stopped having any effect on the structure of the system.

The preceding three methods are unsuitable for detailed analysis
of all the data. Changes in g(r) are small, so the uncertainty
in locating the point at which g(r) stops changing is large.  The other
two methods involve quenches, which are expensive.  A more usable
method,  introduced by  Ullo and Yip,\cite{ullo_yip,ullo_yip2}
determines structural arrest from pressure versus density plots.
At high
pressures, the density of the glass is essentially linear in
pressure,  or vice versa for those who simulate by imposing
a density and measuring the pressure,  as shown in Figure 11.
Ullo and Yip define
the density of the transition point as the intersection
of a straight line through the high-density points  with
a line drawn through the lowest density points.
Even a generous reader would disapprove
if we tried to draw a straight line through our
non-linear low density data, so we will use a slightly different
definition:  $\rho_g$ is the density at which $\rho$ begins to deviate
from the linear behavior of the high-density data.
For the $T^*=0.8$ data, the point at which structural arrest
was determined to occur by other methods is at $P=14,$ and the
density at this point deviates from the value predicted  by the
linear behavior of the higher density points by 1.4\%.   We  therefore
define the structural transition pressure ${P_g}^s$ for  the
squeezing runs by  the
criterion that  it
is the lowest pressure at which density is within
1.5\% of the value extrapolated from the high density linear region.

The values thus obtained for ${P_g}^s,$ with the corresponding
packing fractions ${\eta_g}^s,$  are shown in Table 2.   There is
a small systematic decrease in ${\eta_g}^s$ with  increasing
temperature, which is reasonable  since temperature effects
are included in the WCA calculation.
The structural similarity between all the final states,  at
$P > {P_g}^s,$ as measured by a CNA analysis
of local bonding and by the density after quenching and decompressing,
implies that they are all similar at the point of structural
arrest.\cite{diffusion}
Once a maximally dense type of configuration is attained, barriers to
further structural relaxation in response to pressure apparently diverge
rapidly:  we do not observe any tendency for the higher temperature runs
to achieve denser final states than those squeezed at lower temperature.

If the structure is similar at the point of structural arrest, then
the distribution of barriers to diffusion at this point must be similar.
We have determined this effective energy barrier by plotting
the  diffusion coefficient at the point of structural arrest, ${D_g}^s,$
against the inverse temperature,  as shown in Figure 12.
This clearly shows thermally activated behavior
with a high activation energy, $4.4 \epsilon,$
associated with diffusion near the
point of structural arrest.
The material at this point is densely packed, each atom is
in a cage formed by 12 or so tightly pressed neighbors, and
atomic rearrangement requires crossing a barrier
corresponding to the strength of 4.4 interatomic bonds.

We thus see that a definition of the glass transition as the point at
which further increases
in pressure induce no structural changes is qualitatively  different
from  a definition measuring the point at which diffusion reaches
some constant cutoff, although it is equivalent to a definition
involving diffusion if a thermally activated cutoff is used.
This distinction exists only for the glass transition caused
by increasing pressure: in
the transition caused by cooling,
there is no changing pressure or any other
driving force to induce structural change,  so
all structural rearrangement stops when diffusion stops.
The difference between the two definitions
may not be significant in the laboratory,
since the glass transition is sharp under long time scales, but in
computer simulations,
in which the transition is smeared out almost beyond
recognition (some readers might prefer to leave
out the ``almost''), it is important to distinguish between the two.

The increase with temperature of our
effective packing fraction $\eta_g$ is now understood:
the transition point defined by
a constant cutoff in the diffusion coefficient, although
it is a natural extension of the customary laboratory
definition,  is not well correlated with structural properties,
and $\eta_g$ is a structural property.
At higher temperatures, diffusion continues at a level above the
cutoff until the material  is closer to the final glassy
configuration than at lower temperatures, thus
achieving a higher value for $\eta_g.$

\section{Conclusions}
Of the  structural changes that occur as a two-component LJ
system is compressed at constant temperature,
the most noticeable is a two-fold increase in icosahedral ordering,
as in the transition caused by cooling at constant pressure.
The glass has essentially the same structure whether it was
produced by squeezing at
low or high temperature or by cooling under pressure:
atoms are bonded in the same way, and the volume after quenching
and then releasing the pressure is the same.  Subtle differences
presumably do exist, but their effect on at least these two structural
probes is small.

The effective hard-sphere packing fraction computed using the WCA
method at the pressure $P_g$ for which diffusion falls below a given
cutoff was found to depend on the temperature at which the material
was compressed: it was higher for high  temperature runs.  This
resulted, not from any fundamental difference caused
by squeezing at higher temperature, but rather from the fact that
pressure-driven structural effects are not well correlated
with the diffusion coefficient at any constant time scale.
Pressure-induced structural changes, which for low temperatures
continue to occur well beyond the density
at which diffusion becomes immeasurably small, result in a
denser final structure for
the compressed glass than is achieved by cooling at zero pressure, with a
correspondingly higher effective packing fraction.

The glass transition point defined in terms of a constant cutoff in
diffusion,  analagous to a constant cutoff in viscosity, is
not equivalent to the transition point defined  in
terms of structural properties.
Because diffusion is a thermally
activated process,   the better correlated definition involves an
exponential, thermally activated form for the cutoff in diffusion.
The activation energy thus found to be
associated with diffusion at the point
of structural arrest is large:  we estimate it to be about
$4.4 \; \epsilon.$
\vfill\eject

\section{ Appendix A:  Choosing the Piston Mass}

Andersen's method\cite{HCA} for running isobaric M. D. simulations
couples an extra degree of freedom, the volume of the system,
into the equations of motion.  To do this, an effective
``piston mass'' is needed, which sets the relative strength
of the volume term.    Equilibrium properties
of the system are not actually affected by the piston mass, but dynamical
properties are,  and using an inappropriate value can slow the approach
to equilibrium.
According to Andersen,\cite{HCA} the correct piston mass to use
is such that the
time scale for fluctuation of the volume of the sample
is equivalent to the time scale for sound waves to travel
through the sample.

We show how
to do this here, deriving a simple formula that can be
applied frequently during the simulation to rescale the piston
mass $M_p$ appropriately.

The pressure and volume of the system are functions of the atomic
coordinates.  However, by including an overall scaling factor for
the coordinates, volume can be treated as an independent variable.

The equation of motion for the volume is:
\begin{eqnarray}
{}~~~~~~~~~~~~~~~~~~~~
M_p \ddot V = P - P_{eq}
\end{eqnarray}
where $P_{eq}$ is the externally applied, equibilibrium pressure.
(So if $P  > P_{eq},$  a force acts to  raise
the overall scaling factor and thus the volume,
which reduces the pressure.)
We can Taylor expand V as follows:
\begin{eqnarray}
{}~~~~~~~~~~~~~~~~~~~~
V = V_{eq} + (P - P_{eq}) {dV \over dP}|_{eq} + ...  \cr
{}~~~~~~~~~~~~~~~~~~~~
  = V_{eq} - (P - P_{eq}) \;  \kappa  \; V_{eq} + ...
\end{eqnarray}
where $\kappa$ is the isothermal compressibility $\kappa_{_T}$
for simulations
in an isothermal ensemble (as in this paper), or the adiabatic
compressibility $\kappa_s$ for simulations in a microcanonical ensemble:
\begin{eqnarray}
{}~~~~~~~~~~~~~~~~~~~~
\kappa = -{1 \over V_{eq}} {\partial V_{eq} \over \partial P}
              |_{s \; {\rm or} \; T}.
\end{eqnarray}
Ignoring higher order terms, this gives a differential equation for V:
\begin{eqnarray}
{}~~~~~~~~~~~~~~~~~~~~
M_p \ddot V = -(V - V_{eq})/(\kappa V_{eq}) = -M_p \omega^2 (V - V_{eq})
\end{eqnarray}
where
\begin{eqnarray}
{}~~~~~~~~~~~~~~~~~~~~
\omega \equiv {(M_p \kappa V_{eq})^{-1/2}}.
\end{eqnarray}

We want to choose $M_p$ so that the period $T = 2 \pi/\omega$ coresponds to
the time it takes a sound wave, velocity $v_s,$ to travel the length
of the box, $V_{eq}^{1/3}:$
\begin{eqnarray}
{}~~~~~~~~~~~~~~~~~~~~
2 \pi/\omega = V_{eq}^{1/3}/v_s \Longrightarrow
M_p = [(2 \pi v_s)^2 \kappa {V_{eq}}^{1/3}]^{-1}.
\end{eqnarray}

We now note that $v_s$ is related\cite{tisza} to the adiabatic
compressibility $\kappa_s:$
\begin{eqnarray}
{}~~~~~~~~~~~~~~~~~~~~
{v_s}^2 = 1/(\kappa_s \rho)
\end{eqnarray}

where $\rho$ is the mass density, so  for adiabatic simulations,
\begin{eqnarray}
{}~~~~~~~~~~~~~~~~~~~~
M_p = {\rho \over 4 \pi^2 V_{eq}^{1/3}}.
\end{eqnarray}

For isothermal simulations,  we need to substitute
$\kappa_T =  \gamma \kappa_s,$  where $\gamma =
{c_{_P} / c_{_V}},$  the ratio of the specific
heats.  We have approximated this ratio $\gamma$
as unity in the current work, simply using the same expression
for $M_p$ as would be appropriate for adiabatic
simulations.  This admittedly introduces a systematic underestimation
of $M_p$ by several percent; should a more accurate value for
$M_p$ be required, a more accurate estimate of $\gamma$ should
be used.

\section{ Appendix B:  Effective hard-sphere radius}

We give here a brief sketch of how we computed the effective hard-sphere
radius using the WCA method \cite{WCA} with the corrections of
Verlet and Weiss\cite{VW}.
All of
this is published elsewhere, but we have attempted to sketch the entire
procedure here in one place in a way that enables the reader to quickly
set up a similar calculation using, for example, a mathematical
symbolic manipulation program.

The goal is to find the radius of the hard-sphere system that most closely
resembles a system of particles interacting through some
other kind of potential, in our case, the Lennard-Jones potential.

The WCA method is appropriate for potentials which have a strong
repulsive core.  With this assumption, the approximation is made that
the function
$y(r) = g(r) e^{\beta v_0(r)}$ is similar to that
of a hard-sphere system,  so the radial distribution function
 g(r) may be written as
$g(r) \approx  e^{-\beta v_0(r)} \; y_{_{HS}}(r),$
where $v_0(r)$ is the repulsive
part of
the potential. For the Lennard-Jones system,
\vskip 0.1truein
\begin{eqnarray}
v_0(r) = \cases{ 4 \epsilon [(\sigma/r)^{12} - (\sigma/r)^6]  + \epsilon
                       &~~~~~~~~ $r < 2^{1/6} \sigma$   \cr
                   0   &~~~~~~~~ $r \ge 2^{1/6} \sigma$.  \cr }
\end{eqnarray}
\vskip 0.1truein
Taking the Fourier transform of g(r) to
get the structure factor and equating the structure factor for
the model system to that of a hard-sphere system at $\vec k = 0$
gives\cite{wcavw1}
the WCA formula for the  effective hard-sphere radius of a system
with repulsive potential $v_0(r):$
\vskip 0.1truein
\begin{eqnarray}
\int_0^{\infty} {  y_{_{HS}} \times (1 - e^{-\beta v_{_{HS}}(r)}) \;  r^2 dr}
= \int_0^{\infty} {y_{_{HS}} \times (1 - e^{-\beta v_0(r)}) \; r^2 dr}
\end{eqnarray}
where
\begin{eqnarray}
v_{_{HS}}(r) = \cases{ \infty &~~~~~~~~ $r \le R$ \cr
                          0   &~~~~~~~~ $r > R$. \cr }
\end{eqnarray}

It might seem that just setting the
$\vec k = 0$ peaks equal would be inadequate, that the structure
factor should really be matched as closely as possible
over its entire range, for example
by minimizing the integral over the whole range,
but Verlet and Weiss have performed both calculations
and found effective hard-sphere radii that agree to within a few tenths of
a percent\cite{vwnote}
so we will use the simpler
WCA criterion here.

To get the function $y_{_{HS}}$ needed for calculating R,
we begin with the expression given by Wertheim,\cite{wertheim}
in the form given by
Throop and Bearman\cite{TB} (with typographical errors corrected), for
the radial distribution function g(r) for hard spheres in the Percus-Yevick
approximation.  For hard spheres,  $g(r) = 0$ for $r < R.$
Since R always comes out pretty close to $\sigma,$ certainly always
within a factor of 2, we only need the part for $R < r < 2R$ here, which is:

\begin{eqnarray}
g_{_W}(x,\eta) =
          {1 \over (12 \eta x)} \sum_{1}^{3}{ \lim_{t->t_i} \bigl[ (t - t_i)
            {{t L(t) e^{t (x-1)}} \over {S(t)} } \bigr] }
\end{eqnarray}
\begin{eqnarray}
L(t) &=& 12 \eta [(1 + \eta/2) t + (1 + 2 \eta)]  \cr
S(t) &=& (1 - \eta)^2 t^3 + 6 \eta (1 - \eta) t^2
+ 18 \eta^2 t - 12 \eta (1 + 2 \eta)  \  \cr
x &=& r/R \cr
\eta &=& {\pi \over 6} \rho R^3,
\end{eqnarray}
where $\rho$ is the number density.  The sum is over the three roots of S(t).

This $g(r)$ can be more conveniently written as:

\begin{eqnarray}
g_{_W}(x,\eta) =  \; {1 \over {12 \eta x (1 - \eta)^2}}
\Bigl[
    { {z_1 e^{z_1 (x -1)} L(z_1)} \over {(z_1 - z_2) (z_1 - z_3)}}
  + { {z_2 e^{z_2 (x -1)} L(z_2)} \over {(z_2 - z_1) (z_2 - z_3)}} +
  { {z_3 e^{z_3 (x -1)} L(z_3)} \over {(z_3 - z_1) (z_3 - z_2)}}
\Bigr]
\end{eqnarray}
where $\{z_1,z_2,z_3\}$ are the three roots of $S(t).$

What we need is $y(r) = g(r) - c(r).$  The direct correlation function
$c(r)$ is zero for $r > R$, and Wertheim gives the
following approximation for $r < R:$

\begin{eqnarray}
c_{_W}(r) = \cases{
-\bigl[
(1 + 2 \eta)^2 - 6 \eta (1 + \eta/2)^2 x + \eta/2 (1 + 2 \eta)^2 x^3 \bigr]  /
(1 - \eta)^4
      &$r < R$ \cr
0     &$r \ge R.$ \cr
}
\end{eqnarray}

Verlet and Weiss point out that this $y(r)$ does not match simulation
results
for hard spheres well, but that it can be substantially improved by
a rescaling and the addition of an extra function.   Their improved
$y(r)$ is
\begin{eqnarray}
y_{_{VW}}(r/R,\eta) = g_{_W}(x^{\prime},\eta^{\prime})
                   - c_{_W}(x^{\prime},\eta^{\prime}) + \delta g_1(r,R,\eta),
\end{eqnarray}
with the rescaled variables

\begin{eqnarray}
\eta^\prime &=& \eta (1 - \eta/16)  \cr
x^\prime &=& x (1 - \eta/16)^{-1/3}.
\end{eqnarray}
The extra function
$\delta g_1(r,R,\eta)$ is apparently added over the entire
range $0 < r < 2R.$

\begin{eqnarray}
 \delta g_1(r,R,\eta) = {A \over r} e^{-\mu (r-R)} cos[\mu (r-R)]
\end{eqnarray}
\begin{eqnarray}
A \approx {3 \over 4} R {
    {{\eta^\prime}^2 (1 - 0.7117 \eta^\prime - 0.114 \eta^\prime)^2} \over
    {(1 - \eta^\prime)^4}}
\end{eqnarray}
\begin{eqnarray}
 \mu \approx {{24 A/R^2} \over {\eta^\prime g_{_W}(1,\eta^\prime)}}.
\end{eqnarray}

More accurate, though more complicated,
formulas for $A$ and $\mu$
and also for $c(r)$  have been given by Henderson and
Grundke,\cite{HG}
but we use the simpler Verlet approximations here.

The effective hard-sphere radius, finally, is R for which
equation 2, which may be rewritten this way, is satisfied:

\begin{eqnarray}
\int_0^R{(-c_w(x^{\prime},\eta^{\prime}) + \delta g_1(r,R,\eta)) r^2 dr}
&=&
\int_0^R{(-c_w(x^{\prime},\eta^{\prime}) + \delta g_1(r,R,\eta))
            (1 - e^{-\beta v0(r,\sigma)})r^2 dr}  \cr
&+&
\int_R^{2^{1/6} \sigma}{g_w(x^{\prime},\eta^{\prime}) + \delta g_1(r,R,\eta))
            (1 - e^{-\beta v0(r,\sigma)})r^2 dr}
\end{eqnarray}

We note in passing that while the form for $c(r)$ used here is, according
to Henderson and Grundke,\cite{HG} quite poor,
the final answer for R is quite insensitive to it.  In fact, we
originally computed R with $\delta g_1(r,R,\eta)$ added only for $r > R$,
and the results for R differed from the current results by only a few
tenths of a percent.

\vfill\eject
\acknowledgments This research was  sponsored by the NSF under
grant CHE-9217774.  S.L.S. thanks Arthur P. Smith for many helpful
discussions.
\vskip 0.5truein

$^*$ Present Address: Department of Chemistry, University of British Columbia,
Vancouver, BC, Canada V6T1Y6

\begin{figure}
\caption{
For each temperature simulated, pressure increases over time in a
stepwise fashion, with brief periods of constant, though fluctuating,
pressure alternating with linearly increasing applied pressure.
For one temperature,
$T^* = 0.8,$ we also ran long constant pressure runs from the
endpoint configurations of the short plateaus as illustrated in the inset.
Results from both
are included in data plotted in other  figures.
}
\end{figure}

\begin{figure}
\caption{
Average diffusion constant D, in units of $\sigma_{11}^2/\tau,$ with fitting
functions as described in the text.  The
three different symbols for $T^* = 0.8$ correspond respectively to the
regular short-plateau run of Figure 1, the extended-plateau run of the
inset to Figure 1, and a third run in which the average rate of pressure
increase was slower by about a factor of 8 than in the short-plateau run.
}
\end{figure}

\begin{figure}
\caption{
The transition pressure $P_g$ in the main figure is defined as the pressure
at which the
diffusion constant falls below a cutoff of 0.001.  The inset shows
similar curves for cutoffs ranging from 0.0005 to 0.004.
}
\end{figure}

\begin{figure}
\caption{
  Average effective potential as a function of average interatomic
spacing, with the point corresponding to $P_g$ from Figure 3 indicated
for each temperature.
}
\end{figure}

\begin{figure}
\caption{
The dark atoms illustrated here form a 555 bonded pair:  they
have 5 common neighbors, there are 5 bonds between the common neighbors,
and the longest chain of bonds between the common neighbors consists
of 5 bonds.
An icosahedron is formed of 12 pairs bonded in this way;
one of the dark atoms and each of the light atoms here, paired with
the other dark atom, is one of the 12 pairs, so the atoms
shown here comprise the central atom and half the outer atoms
of a complete icosahedron.  We therefore
refer to 555 pairs as ``icosahedrally bonded.''
}
\end{figure}

\begin{figure}
\caption{
  a.  Each run produces configurations at a sequence of increasing pressures
at a given temperature.  In order to compare the underlying differences
between them, we bring them all into the same state by quenching to
zero temperature, then releasing the pressure. \hfill\break
  b.
We took a quenched state and alternately applied and released a pressure of
$13 \; (\epsilon/\sigma^3).$
Almost no structural change was observed. The number
of 555 pairs is computed based on a cutoff in the maximum distance between
bonded pairs.  We scaled the cutoff by the length scale $V^{1/3}$ in order
to compare configurations in different  pressure states.  Because the
distribution of bond lengths changes slightly under pressure,  this resulted
in a slight difference in the number of pairs found in the compressed states
relative to the uncompressed states, even though no  rearrangements
actually occurred.  This illustrates the subtlety involved in trying
to compare configurations at different pressures,
which is why we generally bring
configurations into the same zero-temperature zero-pressure state before
making comparisons.
}
\end{figure}

\begin{figure}
\caption{
a.  Bonding.  We show here the number of pairs
of several types throughout the glass transition.  The top curve,
showing ``555'' (icosahedrally bonded) pairs,
is the same as in Figure 7b.  The occurrence of
``421'' and ``422'' pairs, associated with FCC and HCP bonding, and
also ``433'' pairs, decreases
gradually throughout the transition, while ``544'' pairs, associated with
distorted icosahedra, increase
slightly. \hfill\break
b. Diffusion and Icosahedral Bonding.  The top curve shows the number of
icosahedrally bonded pairs increasing throughout the glass
transition for a run at $T^*=0.8.$
Each point is an average over the number of 555 pairs in
five different configurations that
had been raised to and held at a pressure P; each configuration was
quenched and decompressed before the structural computation.
 The bottom curves show the diffusion constant, $D_A$ for
small atoms and $D_B$ for large ones, decreasing through the same
transition.
}
\end{figure}

\begin{figure}
\caption{
Diffusion and Isothermal Compressibility.  The compressibility $\kappa$
decreases rapidly in the region where structural changes are occurring to
accommodate the increasing pressure, and then only
 gradually after vitrification,
where pressure continues to force atoms closer together, but without inducing
structural relaxation.  As in Figure 8, $D_A$ and $D_B$ are diffusion constants
for small and large atoms respectively.
}
\end{figure}

\begin{figure}
\caption{
Radial distribution function for component 1
at and beyond $P_g$.   The lower two curves show
the rdf, scaled by the WCA radius,
at the transition point as defined by a constant diffusivity cutoff
for systems compressed at $T^* =0.6$ and at $T^*=1.5.$
The upper
curve is the scaled rdf  for the $T^* =0.6$ run at a much higher
pressure.  Only this latter curve depicts a fully densified
structure: in the lower curves, the split second peak is not
yet fully developed and the more distant peaks have not stopped
moving in relative to the first peak.   The structurally arrested
material of the upper curve undergoes no further changes except for
an overall scaling factor with further increases in pressure, while the
material at the diffusivity-determined $P_g$ has  not yet reached
this point.  Of the two lower curves, the $T^*=1.5$ one is more like
the upper curve; the second and further peaks are slightly closer
together because the material has proceeded farther toward structural
arrest by the time diffusion stops than has the $T^*=0.6$ material.
}
\end{figure}

\begin{figure}
\caption{
Probes of structural arrest.  The upper curve shows the rms difference
between atomic positions before and after quenching as a
function of pressure; the harmonic limit is reached at a pressure of 14.
The lower curve shows the volume after quenching and releasing the
pressure; this stops decreasing at a pressure of 14,  indicating that
the material is fully densified and structurally arrested, so that
further increases in pressure induce no further structural changes.
}
\end{figure}

\begin{figure}
\caption{
Density versus Pressure.  This data is for $T^*=1.5.$ The dark
point, which is identified as the structural transition point,
is where the deviation from the approximately linear behavior of higher
pressures reaches 1.5\%.
}
\end{figure}

\begin{figure}
\caption{
Diffusion Coefficient at the Structural Transition Point.
The natural logarithm of the diffusion coefficient at the
pressure ${D_g}^s$ at which structural arrest occurs, determined as
shown in Figure 11, is well described by a thermally activated
form with an activation energy of $4.4 \; \epsilon.$
Temperature, as usual, is in units of $\epsilon/k_B.$
}
\end{figure}

\begin{table}
\caption{
Effective packing fraction $\eta_g$  near the
point $P_g$ at which diffusion falls below the cutoff.
Values near the analagous point $T_g$ for two cooling runs
are also given for comparison.
Calculation of $\eta_g = (\pi/6) \;  \rho R^3_{_{WCA}}$
requires calculation of an effective radius $R_{_{WCA}}$ for
the particles, which is a function of temperature
and density.
}
\begin{tabular}{cccccc}
Point&
Temperature&
Pressure&
Density&
$\rm R_{WCA}$&
$\eta_g$
\\ \tableline
$\rm P_g$ &0.6  &5.49    &0.908  &1.0270  &0.590\\
$\rm P_g$ &0.7  &7.74    &0.929  &1.0213  &0.594\\
$\rm P_g$ &0.8  &10.47   &0.952  &1.0161  &0.599\\
$\rm P_g$ &1.0  &18.18   &1.007  &1.0070  &0.617\\
$\rm P_g$ &1.5  &42.40   &1.121  &0.9913  &0.655\\
$\rm T_g$ &0.35 & 0      &0.829  &1.0430  &0.564\\
$\rm T_g$ &0.65 & 8      &0.936  &1.0228  &0.601\\
\end{tabular}

\vskip 1.0truein

\caption{
Effective packing fraction ${\eta_g}^s$  near the
point of structural arrest, ${P_g}^s.$   The
systematic trend in the packing fraction  ${\eta_g}^s$
reflects temperature effects included
in the WCA calculation;  other measures indicate that
the structure at the point  of structural arrest is similar
for all the runs, as discussed in the text.
}
\begin{tabular}{ccccc}
Temperature&
Pressure&
Density&
$\rm R_{WCA}$&
${\eta_g}^s$
\\ \tableline
0.6 &10  &0.965  &1.0259  &0.625\\
0.7 &11  &0.968  &1.0206  &0.617\\
0.8 &14  &0.988  &1.0155  &0.620\\
1.0 &15  &0.979  &1.0074  &0.600\\
1.5 &25  &1.021  &0.9908  &0.596\\
\end{tabular}
\end{table}


\begin{references}
\bibitem{average}
The behavior of $D_\alpha$ is the same for both
species, although  about 30-40\% higher for
the smaller atoms;  we could have used $D_1$ instead of $D$, or have
averaged in a different way, without changing any of the
conclusions in this paper.
\bibitem{diffusion} This is certainly true for the three lower temperature
runs, for which diffusion is minimal  beyond the point of
structural arrest.  Some changes could occur beyond ${P_g}^s$ for the
$T=1.5$ run since diffusion continues, but if these involved significant
structural changes, the final configuration should be intrinsically
denser than for the lower temperature runs, which it is not.
\end{references}
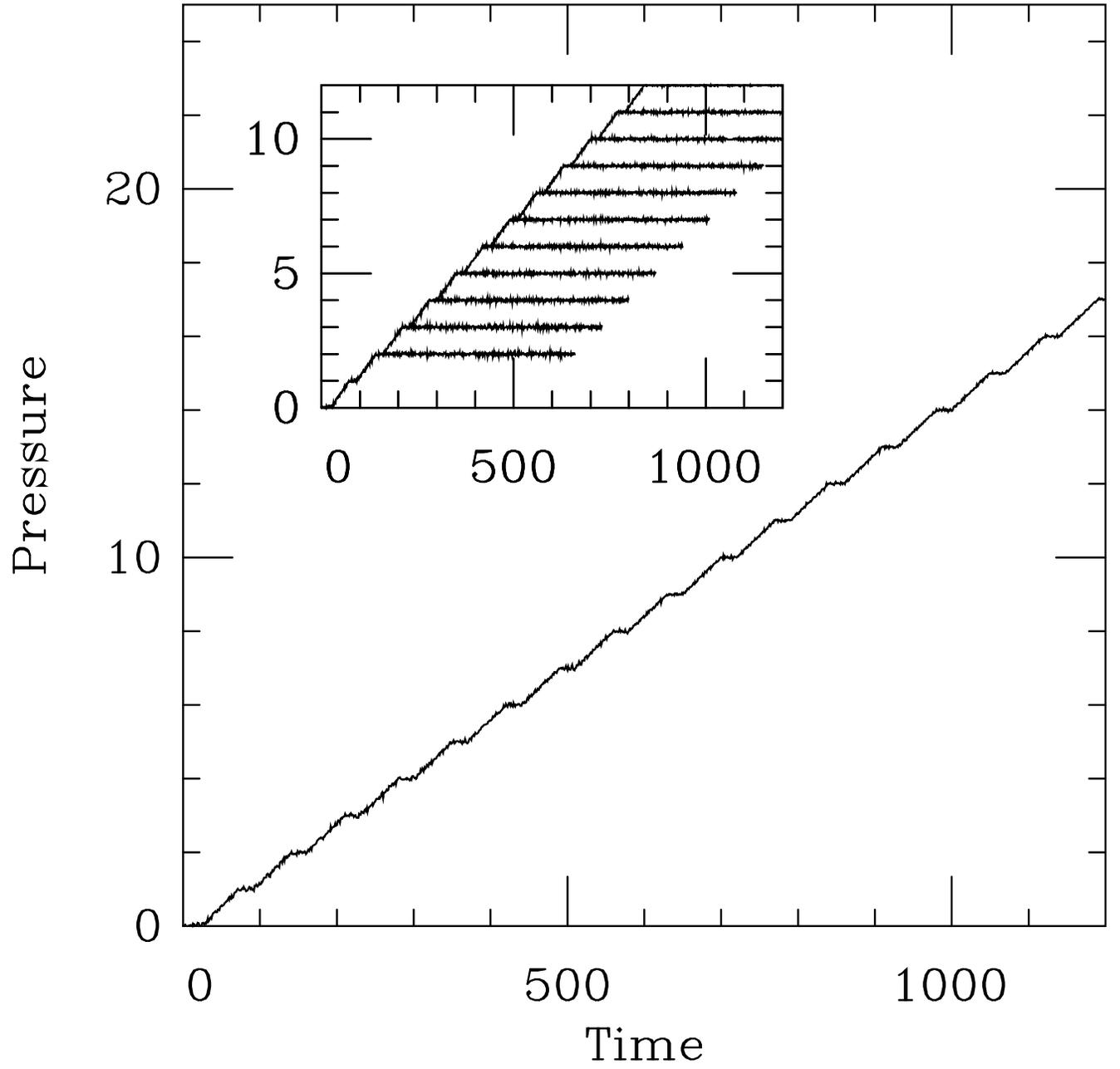
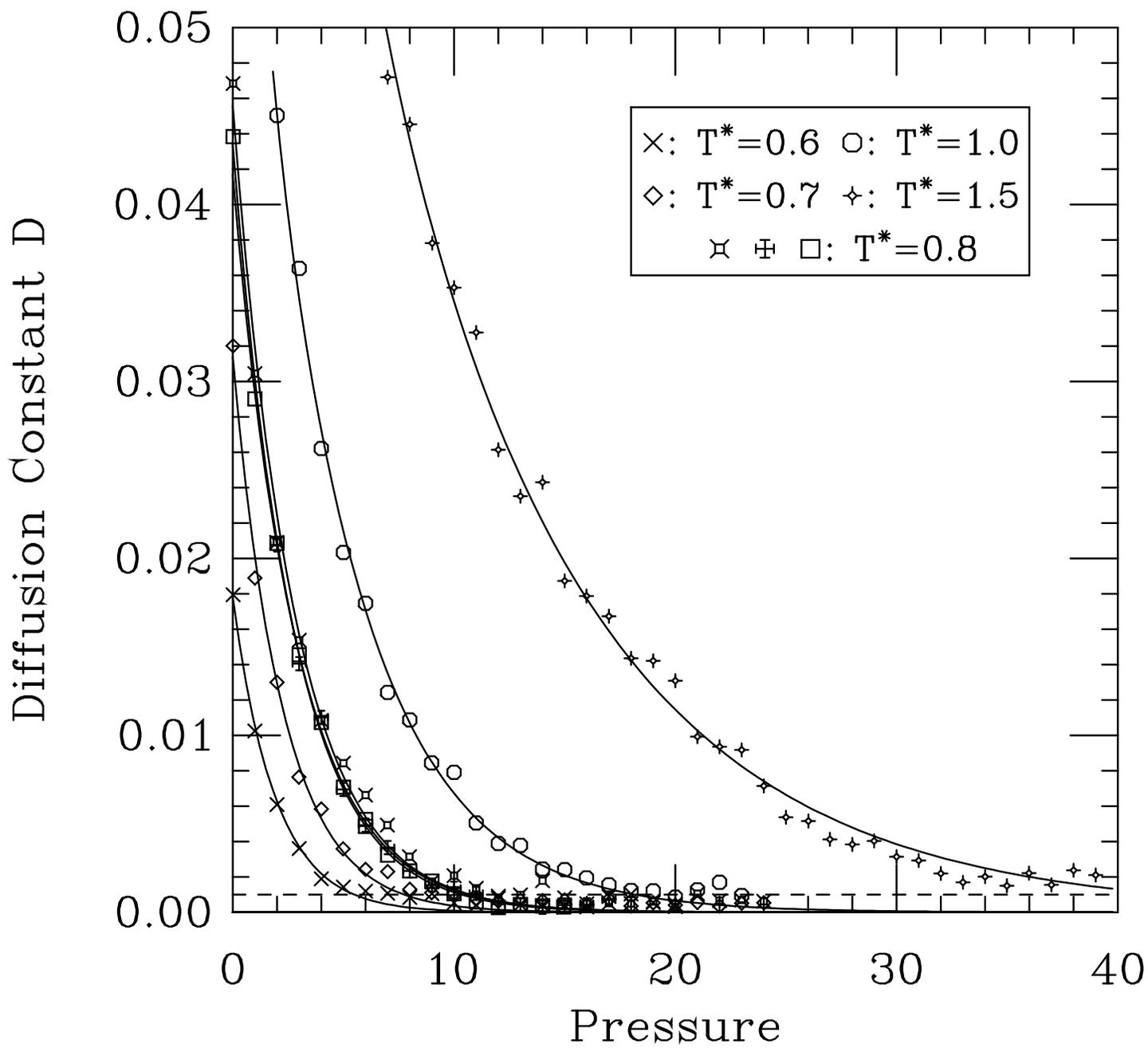
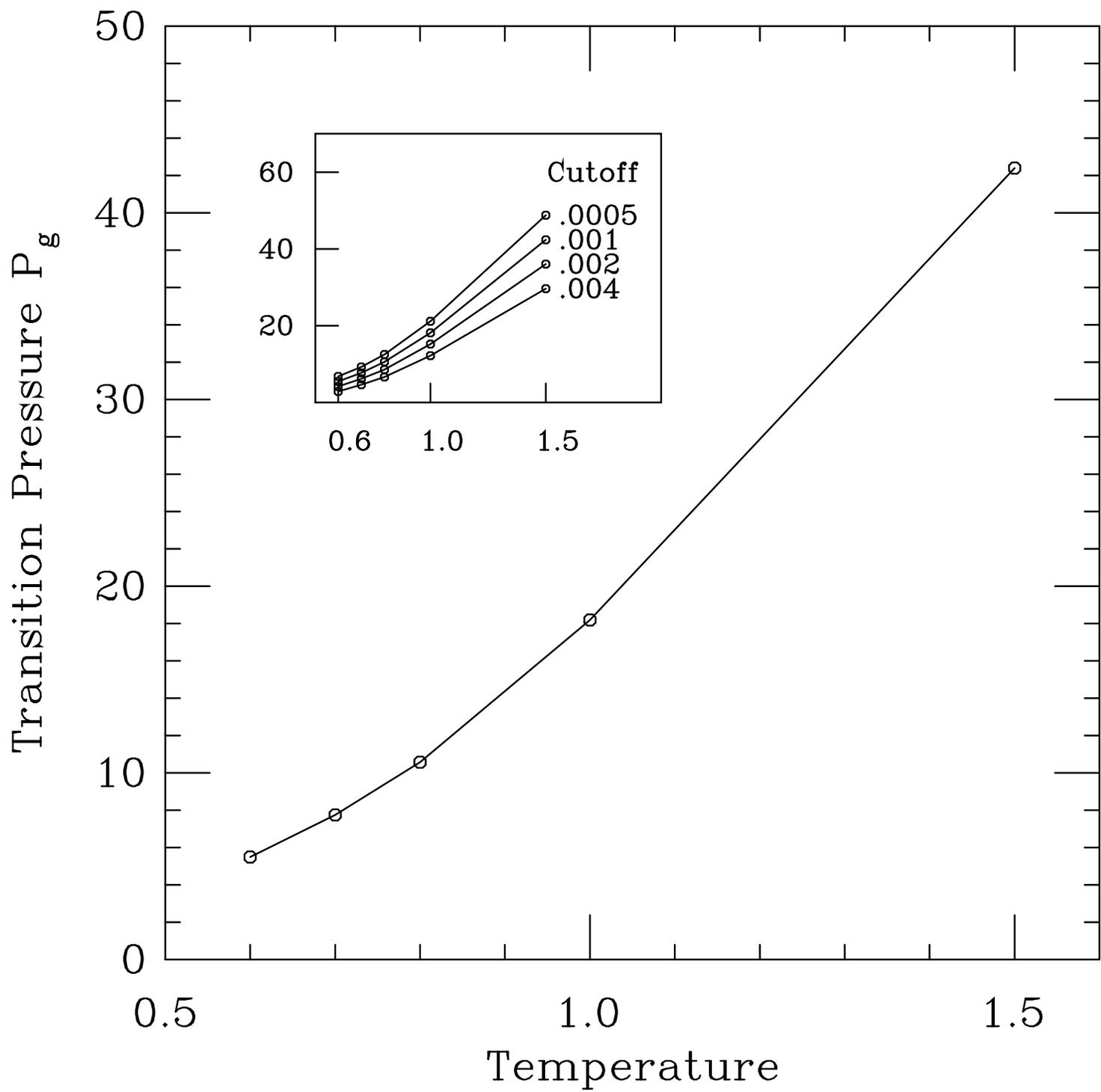
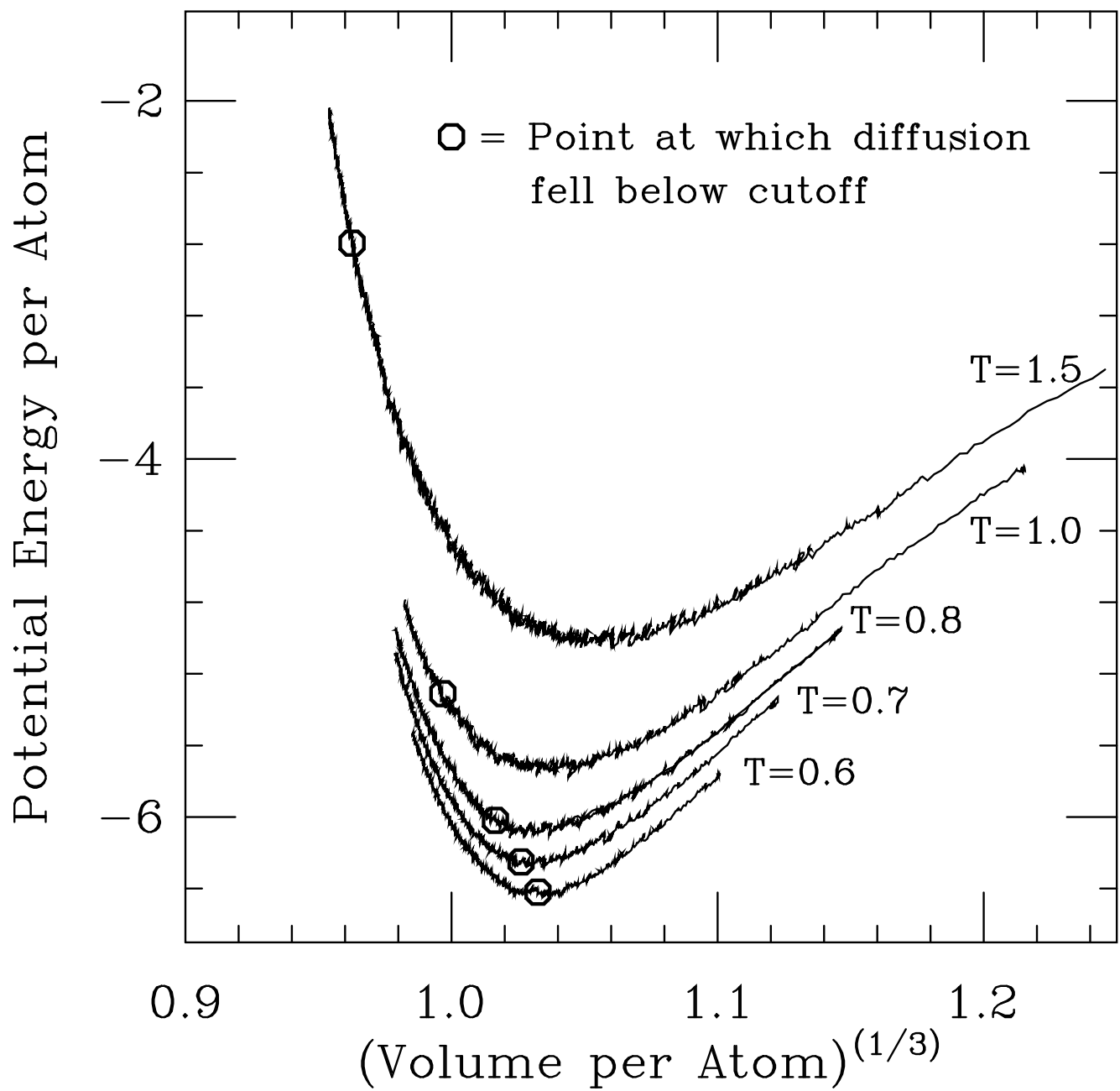
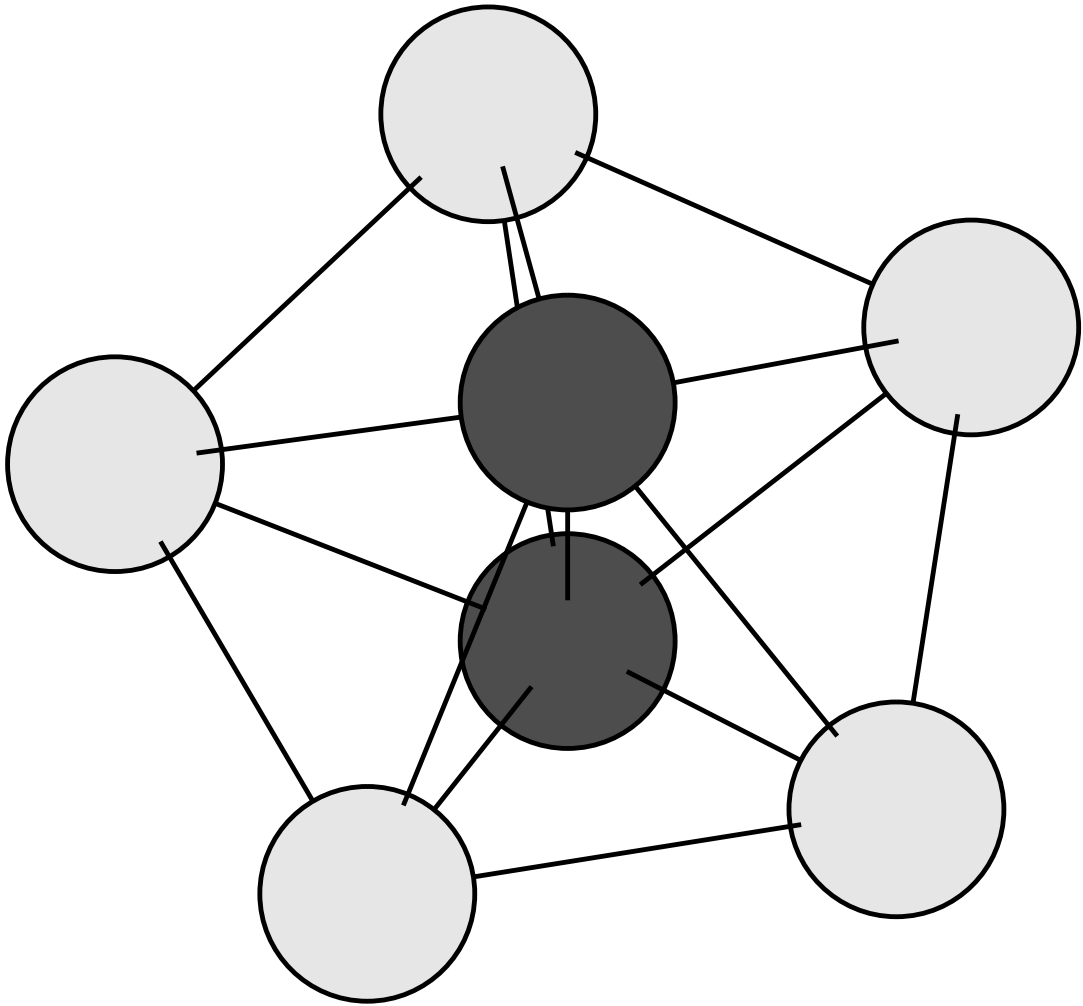
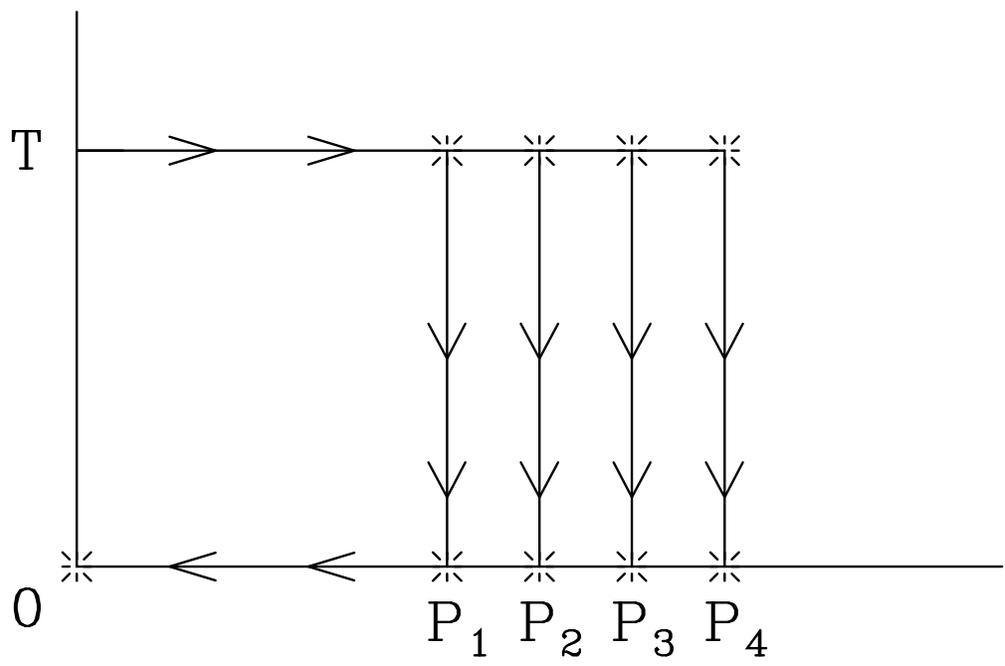
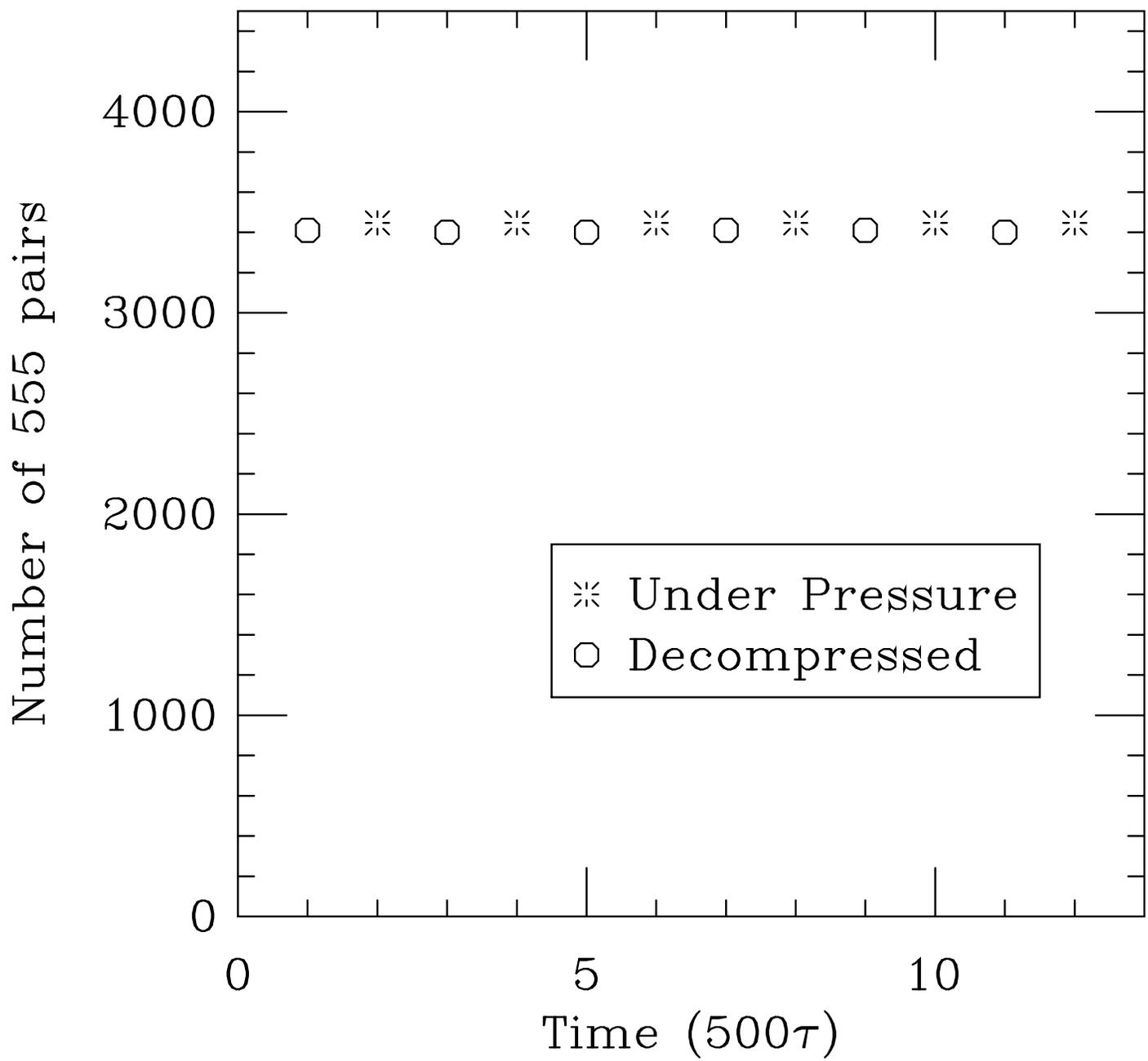
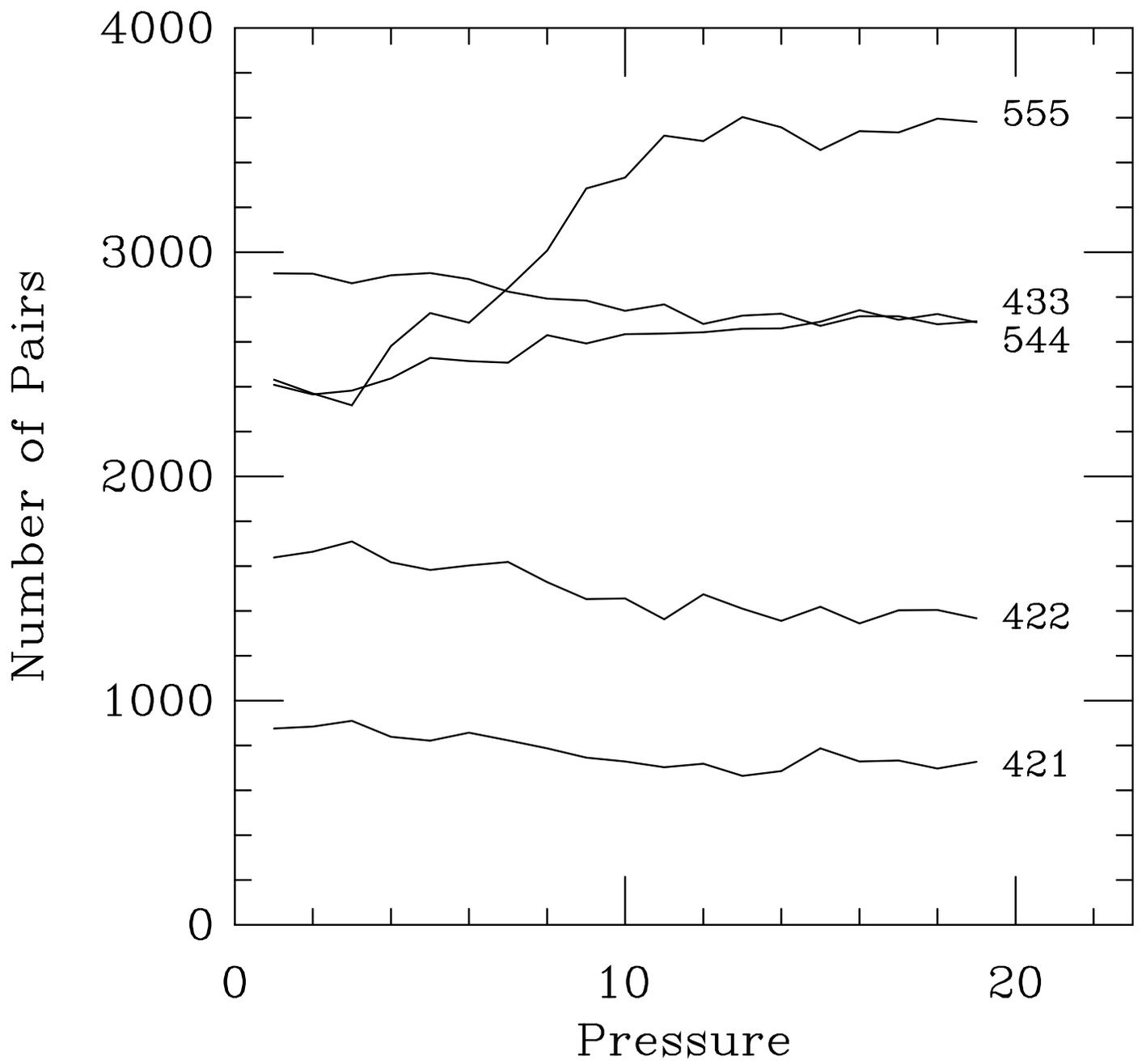
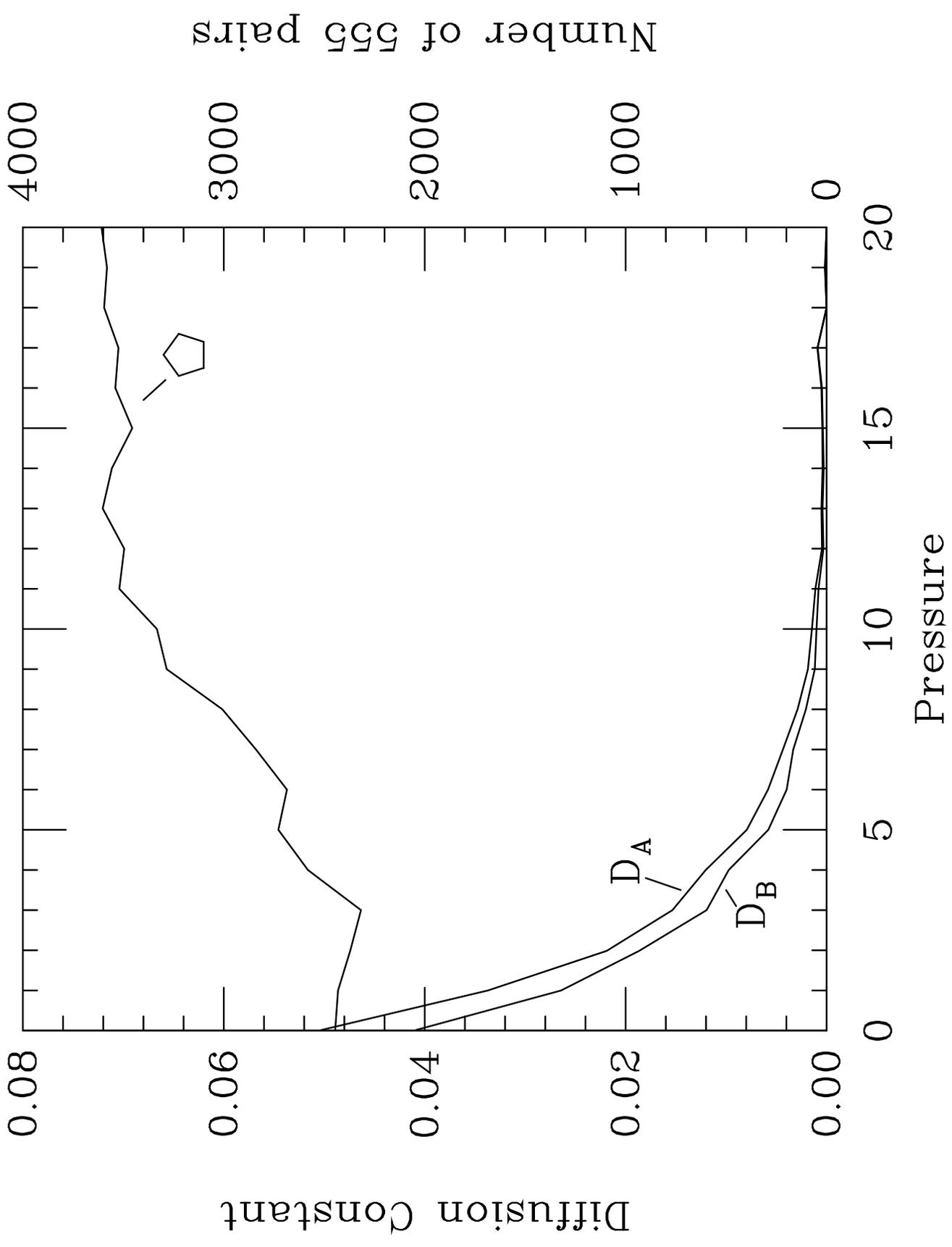
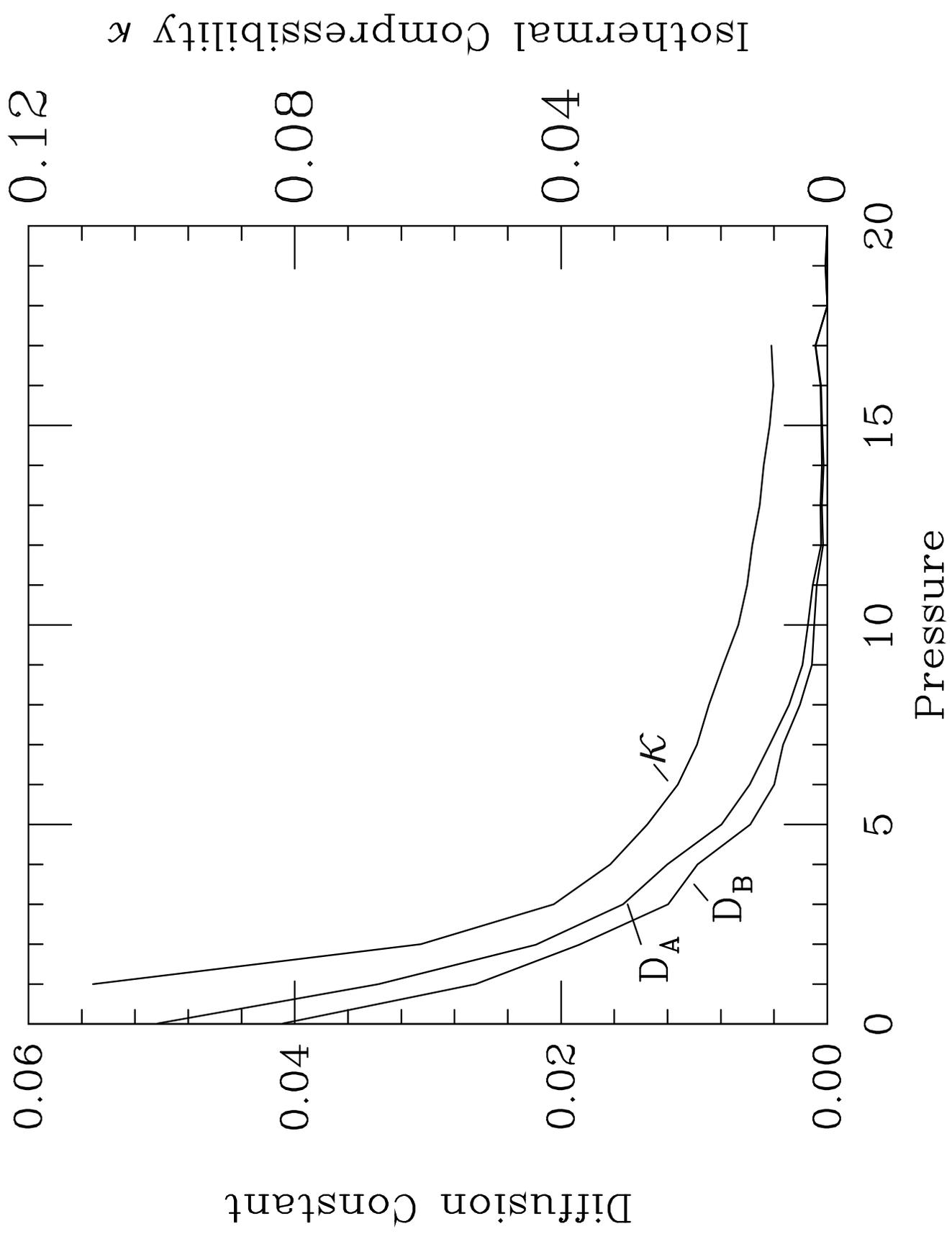
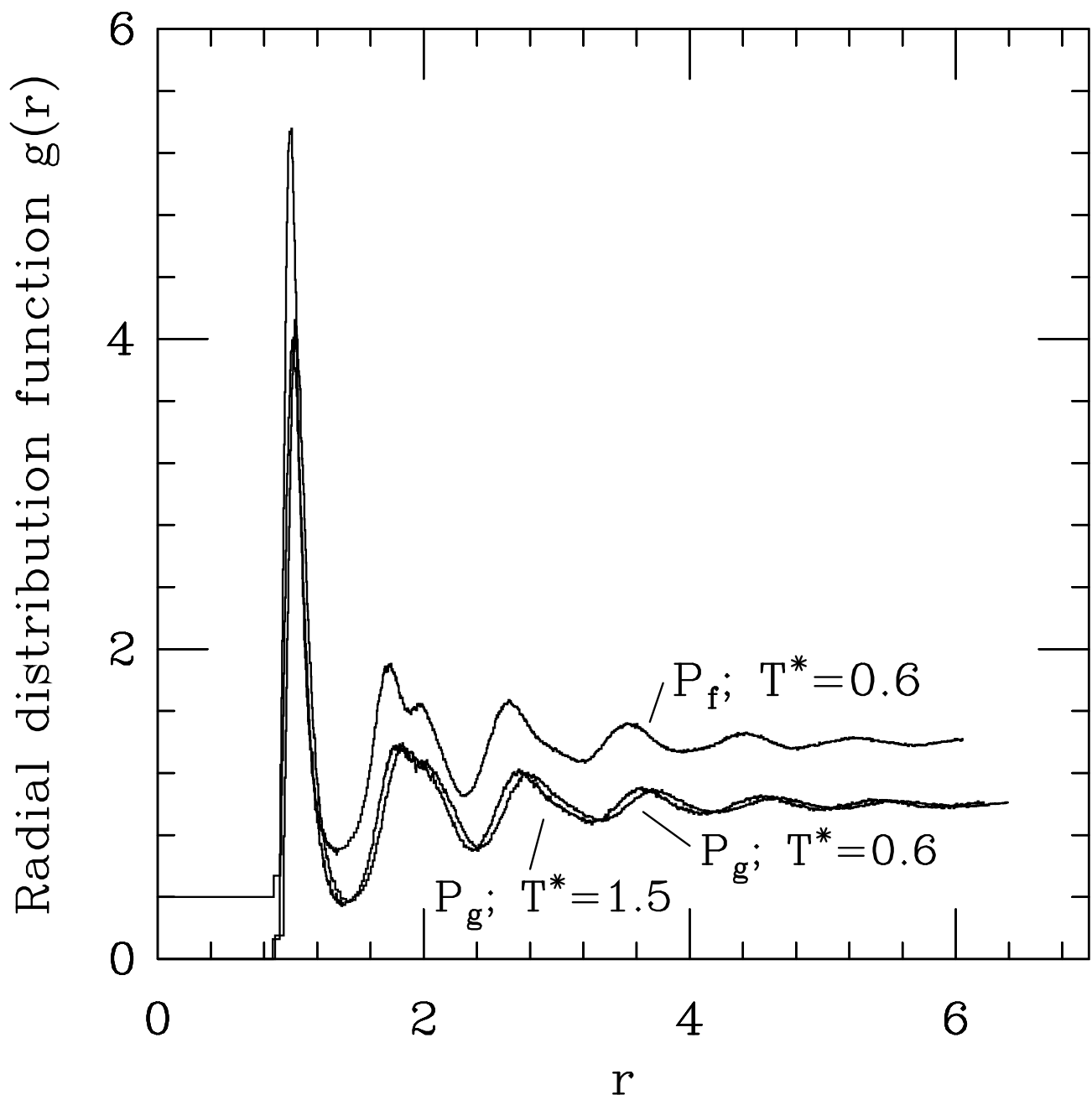
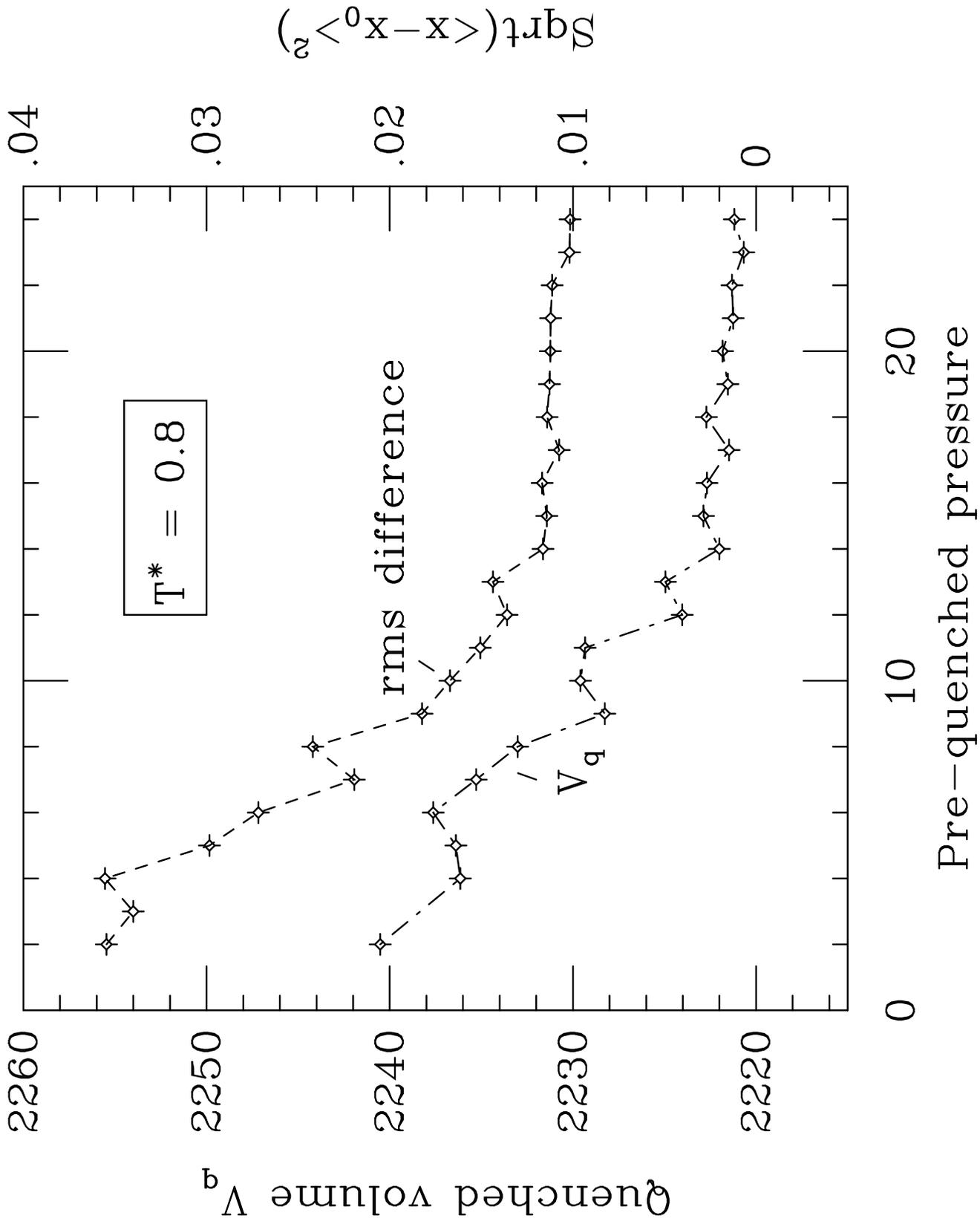
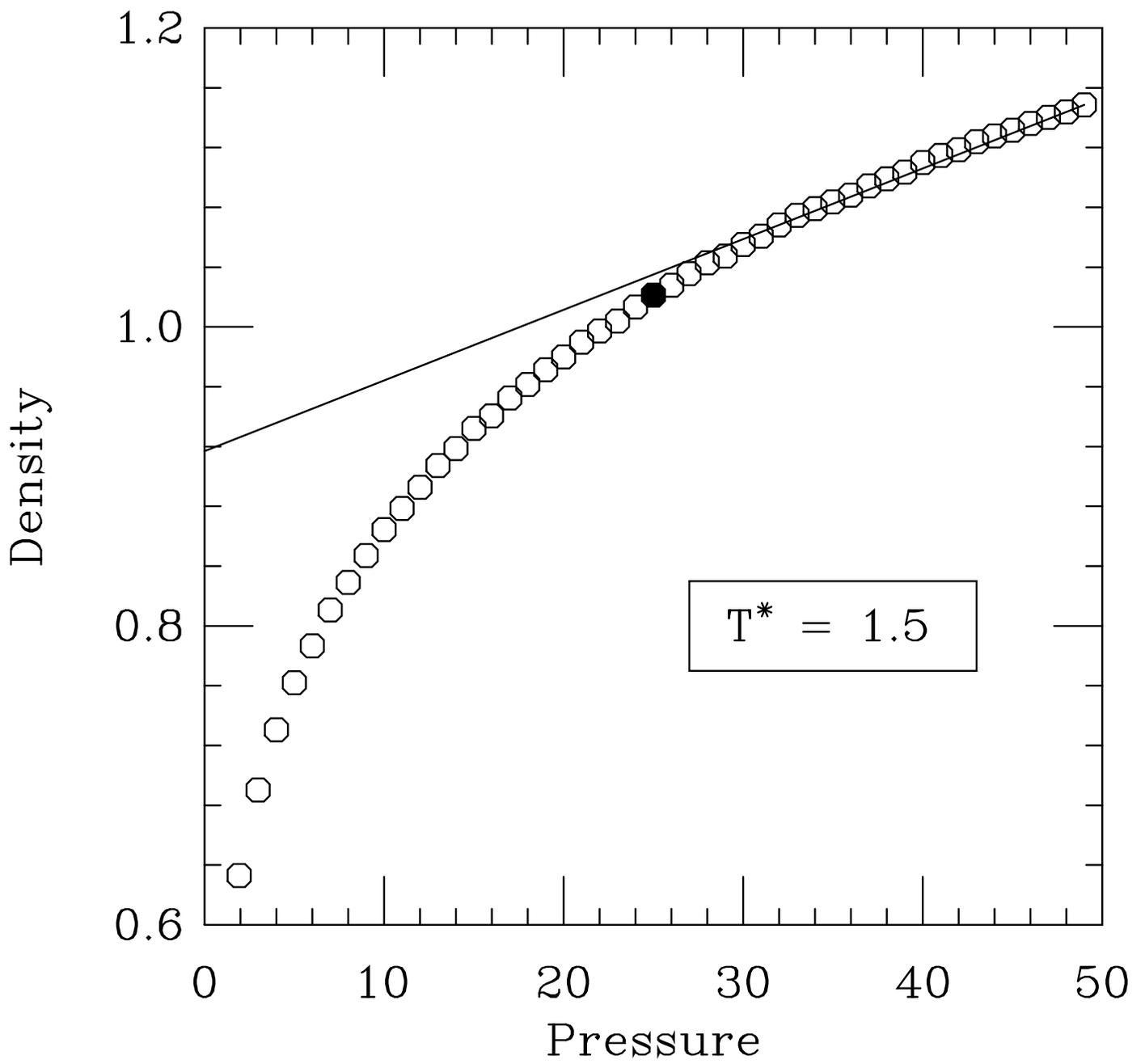
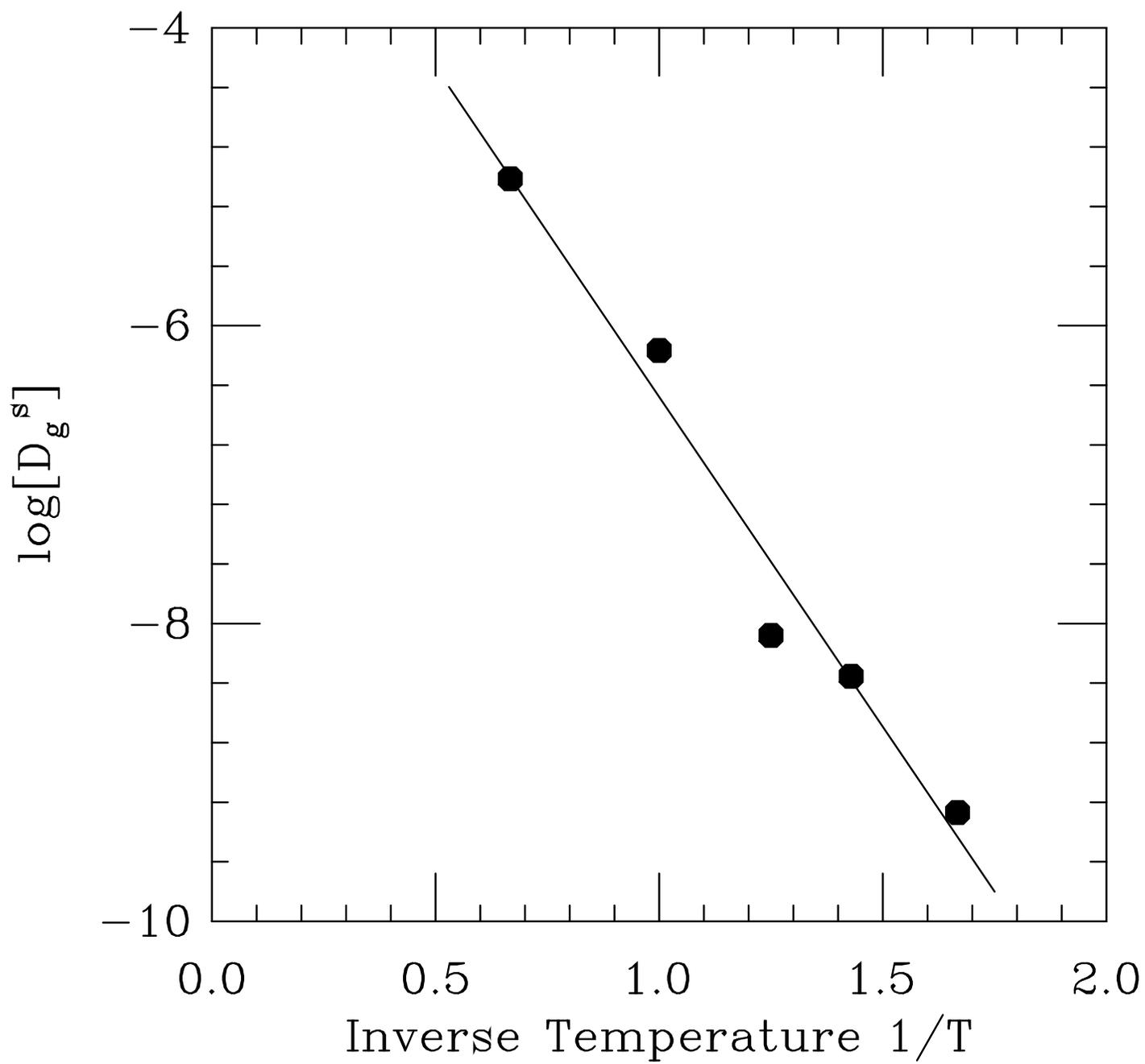
\end{document}